\documentclass[12pt,showpacs,prd]{revtex4}
\usepackage{epsfig,amssymb,amsfonts}
\begin{document}
\title{Vacuum replicas in QCD}
\author{P. J. A. Bicudo}
\affiliation{Grupo Te\'orico de Altas Energias (GTAE), Centro de F\'\i sica das 
Interac\c c\~oes Fundamentais (CFIF),
Departamento de F\'\i sica, Instituto Superior T\'ecnico, Av. Rovisco Pais, P-1049-001 Lisboa, Portugal}
\author{A. V. Nefediev}
\affiliation{Grupo Te\'orico de Altas Energias (GTAE), Centro de F\'\i sica das Interac\c c\~oes 
Fundamentais (CFIF),
Departamento de F\'\i sica, Instituto Superior T\'ecnico, Av. Rovisco Pais, P-1049-001 Lisboa, Portugal}
\affiliation{Institute of Theoretical and Experimental Physics, 117218,\\ B.Cheremushkinskaya 25, 
Moscow, Russia}
\author{J. E. F. T. Ribeiro}
\affiliation{Grupo Te\'orico de Altas Energias (GTAE), Centro de F\'\i sica das 
Interac\c c\~oes Fundamentais (CFIF),
Departamento de F\'\i sica, Instituto Superior T\'ecnico, Av. Rovisco Pais, P-1049-001 Lisboa, Portugal}
\newcommand{\be}{\begin{equation}}
\newcommand{\ee}{\end{equation}}
\newcommand{\ds}{\displaystyle}
\newcommand{\low}[1]{\raisebox{-1mm}{$#1$}}
\newcommand{\loww}[1]{\raisebox{-1.5mm}{$#1$}}
\newcommand{\lmn}{\mathop{\sim}\limits_{n\gg 1}}
\newcommand{\vpint}{\int\makebox[0mm][r]{\bf --\hspace*{0.13cm}}}
\newcommand{\too}{\mathop{\to}\limits_{N_C\to\infty}}
\newcommand{\vp}{\varphi}
\begin{abstract}
The properties of the vacuum are addressed in the two- and four-dimensional quark models for
QCD. It is demonstrated that the two-dimensional QCD ('t~Hooft model) possesses only one
possible vacuum state --- the solution to the mass-gap equation, which provides
spontaneous breaking of the chiral symmetry (SBCS). On the contrary, the four-dimensional theory
with confinement modeled by the linear potential supplied by the Coulomb OGE interaction,
not only has the chirally-noninvariant ground vacuum state, but it possesses an excited
vacuum replica, which also exhibits SBCS and can realize as a metastable intermediate state
of hadronic systems. We discuss the influence of the latter on physical observables as
well as on the possibility to probe the vacuum background fields in QCD.
\end{abstract}
\pacs{12.38.Aw, 12.39.Ki, 12.39.Pn}
\maketitle

\section{Introduction}

Potential models play an important role in studies of QCD. In spite of their obvious 
shortcomings, like loss of Lorentz invariance and causality, they offer a simple and 
intuitive tool for investigation of the chiral properties of the theory, hadronic spectra, 
decays and so on. The Hamiltonian approach, taken together with the Bogoliubov technique for 
diagonalization, provides a powerful method to analyse such potential models. The
two-dimensional 't~Hooft model for QCD, besides being a genuine quantum
filed theory, constitutes an example of an exactly solvable theory 
with an instantaneous linear potential yielded by the two-dimensional
gluon \cite{tHooft}. 

In this paper we address the question whether QCD can possess excited vacuum states, 
hereby called replicas, on top of the usual chirally nonsymmetric vacuum. 
The problem of different vacuum states, their coexistence and stability, 
is not new and was addressed many times in many contexts (as an example, see, {\it e.g.},
\cite{vacuum3} or \cite{LBO} where the \lq\lq history" of the vacuum is briefly outlined). Here we address the 
theoretical possibility of existence of replicas in the framework of quark QCD models which 
have otherwise  been successful in describing hadronic phenomenology and will argue that if one uses  
these sort of models --- which indeed can be traced back to the Gaussian approximation for the cumulant 
expansion of QCD$_4$ --- for their phenomenologic success then one has also to consider the quite possible 
existence of these replicas. They come in the same package. Then it should not 
come as a  surprise that the number of spatial dimensions together with the strength and form 
of the inter-quark potential should play a role of paramount importance in the existence of 
such replicas. We start from the two-dimensional QCD and demonstrate that, for two dimensions, 
only one chirally nonsymmetric state (the vacuum) may exist. Therefore in two dimensions there are 
no replicas. This statement holds true for both, linear and harmonic oscillator
potentials. On the contrary, an infinite set of solutions exists for the mass-gap equation
in four dimensions, if the inter-quark interaction is chosen to be  quadratic in the inter-quark 
distance, as it was found in \cite{ORS2}. In the present paper we confirm this result. 
However, for the more realistic linear potential \cite{linear} the situation in QCD$_4$ changes dramatically. 
We study the corresponding mass-gap equation and find the {\em pure} linearly rising interaction 
to fail {\em just} to be sufficiently \lq\lq binding" to hold any replicas,  
so that only one chirally nonsymmetric solution to the mass-gap equation may exist 
in this case (the trivial chirally symmetric solution is always present in 3+1, 
as opposed to the two-dimensional case). 
On the contrary, for the more physical interaction with realistic values, both for the string
tension $\sigma_0$ and the strong coupling constant $\alpha_s$ of the Coulomb interaction 
together with a constant term $\gamma_\mu U \gamma_\mu $ , only to adjust for the right 
chiral condensate, we find one vacuum replica to exist besides the usual vacuum state. 
We give qualitative
arguments that the excited vacuum state should exist in the real QCD and discuss its
possible influence on physical observables.

This paper is organized as follows. In the second section we consider the two-dimensional
't~Hooft model and briefly review the formalism going through the concepts
of the Hamiltonian approach, the Bogoliubov transformation from bare to dressed quarks,
and the mass-gap equation for the chiral angle. We argue that the mass-gap equation for
the 't~Hooft model has only one solution, found numerically in \cite{Ming Li}, which
defines the chirally noninvariant phase of the theory, whereas the phase with unbroken chiral
symmetry possesses an infinite energy and, hence, is unphysical. These results are found
to remain for a similar, but technically simpler, situation when using the harmonic 
oscillator instead of the linear potential. In the third section we
deal with the case of the four-dimensional QCD and study the mass-gap equation
numerically, both for harmonic and linear confining quark kernels. We find that an enumerable infinite set of 
replicas always exists in the case of the quadratic potential, whereas the linear potential  
appears to be just not quite enough \lq\lq binding\rq\rq, so that it has to be added either to a Coulomb
or a constant potential (or to  a combination of both) to support the existence of 
an excited vacuum state. It turns out that for a realistic set of parameters 
$\{\sigma_0,\;\alpha_s,\;U\}$, 
simultaneously describing hadronic spectra {\em and} yielding a quark condensate in the ball-park of
$\langle {\bar q} q\rangle=-(250MeV)^3$ it was {\em impossible } to avoid the existence of one replica.
The fourth section is devoted to the discussion of the hadronic processes in the presence of the
excited vacuum. We give arguments on how the excited vacuum replica can affect the
physical observables and serve as a probe of the QCD vacuum structure.
Our conclusions and the outlook are the subject of the fifth section. 
Throughout this paper we consider the chiral limit, {\it
i.e.}, we always put the mass of the quark $m$ equal to zero from the very beginning.

\section{Two-dimensional QCD ('t~Hooft model)}

\subsection{Introduction to the model}

The two-dimensional 't~Hooft model, suggested in 1974 \cite{tHooft} and widely 
discussed in literature as a toy model for QCD, is described by the Lagrangian
density
\be
L (x)= -\frac{1}{4}F^a_{\mu\nu}(x)F^a_{\mu\nu}(x) + \bar q(x)i\hat{D}q(x),
\label{lagrangian}
\ee
$$
\hat{D} = (\partial_{\mu} - igA^a_{\mu}t^a)\gamma_{\mu}
$$
and the large-$N_C$ limit implies that \footnote{We consider the weak limit of
the theory $m\gg g\sim 1/\sqrt{N_C}$, so that the limit $N_C\to\infty$ is to be
taken prior to $m\to 0$ \cite{Zhitnitsky}.}
\be
\gamma=\frac{g^2N_C}{4\pi}\too {\rm const}.
\ee

We fix the axial gauge imposing the condition $A_1(x_0,x)=0$
\cite{Bars&Green,we1}, so that the vector-field propagator takes the form
\be
D^{ab}_{\mu\nu}(x_0 - y_0, x - y ) =-\frac{i}{2}\delta^{ab}g_{\mu 0}g_{\nu 0}
|x-y|\delta(x_0-y_0),
\label{D}
\ee
and integrate out
gluonic degrees of freedom to arrive at the Hamiltonian,
$$
H=\int dxq^{+}(x)\left(-i\gamma^5\frac{\partial}{\partial x}\right) 
q(x)-\hspace*{3cm}
$$
\be
\hspace*{3cm}
-\frac{g^2}{2}\int dx\int
dy\;q^{+}(x)\frac{\lambda^a}{2}q(x)q^{+}(y)\frac{\lambda^a}{2}q(y)\frac{\left|x-y\right|}{2}.
\label{H}
\ee

Following the standard BCS technique, we introduce the dressed quark field
\cite{Bars&Green},
\be
q_{\alpha}(t,x)=\int\frac{dp}{2\pi}e^{ipx}[b_{\alpha}(p,t)u(p)+d_{\alpha}^+
(-p,t)v(-p)],
\label{quark_field}
\ee
\be
b_{\alpha}(p)|0\rangle= d_{\alpha}(-p)|0\rangle =0,\quad
b^{+}_{\alpha}(p)|0\rangle=|q\rangle,\quad d^{+}_{\alpha}(-p)|0\rangle=|\bar{q}\rangle,
\label{bnd}
\ee
\be
\ds\{b_{\alpha}(p,t)b^+_{\beta}(q,t)\}=
\ds\{d_{\alpha}(-p,t)d^+_{\beta}(-q,t)\}=2\pi\delta(p-q)\delta_{\alpha\beta},
\label{bdcommutators}
\ee
\be
u(p)=T(p)\left(1 \atop 0 \right),\quad v(-p)=T(p)\left(0 \atop 1 \right),
\label{unv}
\ee
$$
T(p)=e^{-\frac{1}{2}\theta(k)\gamma_1},\quad-\frac{\pi}{2}\leq\theta\leq\frac{\pi}{2},
\quad\theta(-p)=-\theta(p),
$$
where the Bogoliubov-Valatin angle $\theta$ is subject to mass-gap equation. In the
new basis the Hamiltonian (\ref{H}) takes the form:
\be
H= LN_C{\cal E}_v + :H_2: + :H_4:,
\label{Hh}
\ee
where $LN_C{\cal E}_v$ stands for the vacuum energy ($L$ being the one-dimensional volume)
and the second and the third terms on the r.h.s. for the  quadratic and
quartic in the quark/antiquark
creation/annihilation operators, respectively. In the remainder of this paper 
we shall concentrate on the energy density ${\cal E}_v$.

\subsection{The vacuum energy and the chiral symmetry breaking}

The vacuum energy $LN_C{\cal E}_v$ can be calculated as an average of
the time ordered Hamiltonian (\ref{H}) over the BCS vacuum state 
\footnote{We put tilde in the notation of the BCS vacuum, 
$|{\tilde 0}\rangle$, unnihilated by ${\tilde b}$ and ${\tilde d}$, 
to distinguish it from the trivial one, $|0\rangle$, annihilated by
the bare operators $b_0$ and $d_0$.},
\be\label{vacEN}
E_{\rm vac}=\langle {\tilde 0}|TH({\tilde b},{\tilde b^+},{\tilde d},{\tilde d^+})|{\tilde 0}\rangle\equiv LN_C{\cal E}_{\rm vac}=
LN_C({\cal E}_{\rm vac}^{\rm free}+
\Delta{\cal E}_{\rm vac}),
\ee
where we have introduced the 
excess of the vacuum energy density over the free-theory energy density,
${\cal E}_{\rm vac}^{\rm free}$. We have
\be
\Delta{\cal E}_{\rm vac} =\int\frac{dp}{2\pi}Tr
\left\{\gamma^5p\Lambda_{-}(p)+\frac{\gamma}{4\pi}
\int\frac{dk}{(p-k)^2}\Lambda_{+}(k)\Lambda_{-}(p)\right\}
-{\cal E}_{\rm vac}^{\rm free},
\label{vac}
\ee
$$
\Lambda_{\pm}(p)=T(p)\frac{1\pm\gamma^0}{2}T^+(p)
$$
or, in terms of the angle $\theta$,
\be
\Delta{\cal E}_{\rm vac}[\theta]=-\int\frac{dp}{2\pi}(p\sin\theta(p)-|p|)-\frac{\gamma}{4\pi}
\int\frac{dpdk}{(p-k)^2}\cos\left[\theta(p)-\theta(k)\right],
\label{evac}
\ee
where for the free massless theory we substituted $\gamma=0$ and 
$\theta_{\rm free}(p)=\frac{\pi}{2}{\rm sign}(p)$.

The excess $\Delta{\cal E}_{\rm vac}[\theta]$  is the main object of our study. The true angle $\theta$
should minimize the vacuum energy, that is, it is a solution to
the mass-gap equation,
\be
\frac{\delta \Delta{\cal E}_{\rm vac}[\theta]}{\delta\theta(p)}=0.
\label{massgap}
\ee 

First, we perform a simple qualitative analysis of Eq.(\ref{evac}). 
Let $\theta(p)$ be the solution of (\ref{massgap}) corresponding to the minimum of
$\Delta {\cal E}_{\rm vac}[\theta]$. Then for $\theta(p/A)$, with  an arbitrary parameter
$A\ne 1$, $\Delta {\cal E}_{\rm vac}[\theta /A]> \Delta {\cal E}_{\rm vac}[\theta]$. 
Naive dimensional analysis shows that $\Delta {\cal E}_{\rm vac}$ scales 
with $A$ as
\be\label{erroneous}
\Delta{\cal E}_{\rm vac}=\frac12C_1A^2-\gamma C_2
\label{EA}
\ee
with $C_{1,2}$ being positive constants. Then the stable solution is provided by
 minimizing the energy (\ref{EA}) with respect to $A$, that is, by $A_0=0$, 
which corresponds either to the free massless theory,
\be
\theta_{\rm free}(p)=\frac{\pi}{2}{\rm sign}(p),\quad E_{\rm free}(p)=|p|,
\label{frees}
\ee
with $E_{\rm free}(p)$ being the free-quark dispersive law, or
to the analytical solution of the mass-gap equation found in \cite{Bars&Green}, which
reads:
\be
\theta(p)=\frac{\pi}{2}{\rm sign}(p),\quad E(p)=|p|-P\frac{\gamma}{|p|},
\label{BGs}
\ee
where the symbol $P$ stands for the principal-value prescription. Both solutions
yield unbroken chiral symmetry. Thus one would have arrived at the 
erroneous  conclusion that no chirally nonsymmetric solution of the mass-gap equation may exist, 
were not for the fact that the vacuum energy (\ref{evac}) is logarithmically infrared divergent 
and the correct form of the relation (\ref{EA}) should have been
\be
\Delta{\cal E}_{\rm vac}=\frac12C_1A^2-\gamma C_2\ln A+\gamma C_3
\label{EA2}
\ee
instead of Eq.(\ref{erroneous}), with $C_3$ proportional to the logarithm of the 
cut-off \footnote{Notice that there is no way to get rid of
the logarithmic divergence in $\Delta {\cal E}_{\rm vac}$ defined by Eq.(\ref{evac}). If
the infrared behavior of the integrand is improved, for example, by
changing $\cos\left[\theta(p)-\theta(k)\right]$ for
$\cos\left[\theta(p)-\theta(k)\right]-1$ (principal-value prescription), then the ultraviolet logarithmic
divergence appears. Thus one cannot
remove both divergences in (\ref{evac}) simultaneously.}. The function (\ref{EA2}) always has a 
minimum at
\be
A_0=\sqrt{\gamma\frac{C_2}{C_1}}\neq 0,
\label{A0}
\ee
which corresponds to a nontrivial solution of the mass-gap equation, found
numerically in \cite{Ming Li}. From the form of the function (\ref{EA2}) one
can see the logarithmic growth of the energy when approaching the solution
(\ref{BGs}),
which corresponds to $A_0=0$ \footnote{Once the logarithmically divergent term in 
(\ref{EA2}) is proportional to the coupling constant $\gamma$, then the
theory still possesses the free limit (\ref{frees}) (which also corresponds to
$A_0=0$), when $\gamma$ tends to zero and the
logarithmic term disappears.}. Therefore two conclusions can be deduced. First, 
there is no need
to demand  $\Delta{\cal E}_{\rm vac}$ to be negative as the nontrivial vacuum
energy defined by (\ref{A0}) is always compared with the infinite energy of the chirally 
symmetric
phase. Second, no phase transition of chiral symmetry restoration is possible
in the 't~Hooft model, its phase diagram being trivial. A similar conclusion is made in
\cite{G}.

The chiral condensate $\langle\bar qq\rangle$ is given by
\be
\langle\bar qq\rangle=-\frac{N_C}{\pi}\int_0^{\infty}dp\cos\theta(p).
\label{cc}
\ee
Then the nontrivial solution, found in \cite{Ming Li}, provides a nonvanishing chiral
condensate, whereas the solution (\ref{BGs}) gives $\langle\bar qq\rangle=0$, 
but with an {\em infinite} energy. Thus we conclude that in the 't~Hooft model 
there can only be one phase,
with spontaneously broken chiral symmetry. Note that there is no contradiction
to the Coleman theorem \cite{Coleman}, which precludes spontaneous breaking of
symmetries in two-dimensional theories, if one considers infinite-$N_C$ limit
\cite{Witten,Zhitnitsky}. 

The contribution of the infrared logarithmic divergence in (\ref{EA2})
has yet another consequence for the 't~Hooft model --- namely, that the nonvanishing
chiral condensate and the spontaneous chiral symmetry breaking happen already
at the BCS level, whereas going beyond BCS just provides corrections suppressed by
powers of $N_C$. Indeed, staying at the BCS level, one diagonalizes the theory in
the one-particle sector, introducing a nontrivial vacuum, full of correlated $q\bar q$ pairs.
It was demonstrated in \cite{we1} that one can go beyond BCS in the 't~Hooft model
performing the second Bogoliubov-like transformation and diagonalizing the theory in
the mesonic sector, that is, in the sector of the quark-antiquark bound states.
The true vacuum of the theory, $|\Omega\rangle$, is connected to the BCS vacuum,
$|{\tilde 0}\rangle$, via a unitary operator. One should use the true vacuum
state for calculations of all matrix elements of operators, like the chiral condensate, but as shown in
\cite{we1}, the difference between the two averages is of the next-to-leading order
in $N_C$:
\be
\langle \Omega|\bar qq|\Omega\rangle=\langle {\tilde 0}|\bar
qq|{\tilde 0}\rangle+O\left(\sqrt{N_C}\right)\sim O(N_C)+O\left(\sqrt{N_C}\right).
\label{Om}
\ee

It is instructive to compare the 't~Hooft model with other two-dimensional
models, {\it e.g.}, with the Gross-Neveu one \cite{GN}. The latter is given by the interaction
Lagrangian
\be
L_{int}=\frac12g\left(\bar\psi_i\psi_i\right)^2,\quad g>0,
\label{GNm}
\ee
where the flavour index $i$ runs from 1 to $N\gg 1$, whereas $\lambda=g^2N$ remains
finite. The positive sign of the constant $g$ chosen in (\ref{GNm}) is known to lead to condensation
in the singlet $\bar\psi\psi$ channel \cite{GN}, whereas the negative sign leads to 
charged $\psi\psi$ and $\bar\psi\bar\psi$ condensates \cite{Remb}.
It was shown that the chiral symmetry is also broken in the Gross-Neveu
model, but in order to have a nontrivial chiral condensate one has to go beyond BCS level,
summing fermionic bubbles. Bosonization of the model, in terms of the compound
state $\sigma=g\bar\psi\psi$, appears to be the most convenient formalism for
doing this and the resulting renormalized \lq\lq potential" reads:
\be
V(\sigma,\sigma_0)=\frac12\sigma^2+\frac{\lambda}{4\pi}\sigma^2
\left[\ln\left(\frac{\sigma}{\sigma_0}\right)^2-3\right]
\label{V}
\ee
with $\sigma_0$ being the renormalization point, which brings the mass 
scale into the theory. Only the first term in (\ref{V}) appears at the BCS
level leading to the trivial solution $\sigma_{\rm min}=0$. On the other hand, the 
full potential (\ref{V}) provides also a nontrivial solution,
\be
\sigma_{\rm min}=\pm\sigma_0\exp\left(1-\frac{\pi}{\lambda}\right).
\ee

From Eq.(\ref{V}) one can see that the logarithmic term, which scales as $\ln A$
under the above mentioned transformation, is
multiplied by $\sigma^2\sim A^2$ which cancels the logarithmic growth of the potential
at the origin. Thus the two phases, chirally symmetric and
nonsymmetric, coexist in the model, whereas the latter, with nonvanishing $\sigma_{\rm min}$, is
energetically preferable. 

The chiral condensate of the Gross-Neveu model scales like
\be
\langle\sigma\rangle\sim \langle g\bar\psi\psi\rangle\sim gN\sim\lambda\sqrt{N}\sim\sqrt{N}
\label{sig}
\ee
for large $N$. Relation (\ref{sig}) is analogue of (\ref{Om}) with
vanishing leading term of order $O(N)$, which is
in agreement with the statement made above, that the chiral condensate in the
Gross-Neveu model vanishes at the BCS level. One can arrive at the same
conclusion from simple dimensional considerations. Indeed, the Gross-Neveu model
describes scale-invariant renormalizable theory, so that there is no parameter
with dimension of mass in the bare Lagrangian and one has to deal with the mechanism of
the dimensional transmutation to get a scale. In contrast, the 't~Hooft
Lagrangian (\ref{lagrangian}) defines a super-renormalizable theory and the
dimensional parameter, the coupling constant $g$, is present in the
theory from the very beginning.

Formula (\ref{EA2}) can be given a more physically transparent form if one notices that
the chiral condensate $\Sigma\equiv\langle\bar{q}q\rangle$ from Eq.(\ref{cc}) 
scales linearly with $A$, 
$\Sigma \to A \Sigma$, so that it can be used instead of $A$,
\be
\Delta{\cal E}_{\rm vac}=C_1'\left[\frac12\left(\frac{\Sigma_{\phantom 0}}{\Sigma_0}\right)^2-
\ln\left|\frac{\Sigma_{\phantom 0}}{\Sigma_0}\right| \right]+\gamma C_3',
\label{EA3}
\ee
where $\Sigma_0$ stands for the actual value of the chiral condensate. 

\subsection{Harmonic oscillator-type interaction}
In this subsection we complete the study of two dimensional Hamiltonians 
using a harmonic oscillator-kernel in (\ref{H}), instead of the linear kernel,
\be
\gamma|x-y|\to K_0^3(x-y)^2,
\ee
or, in momentum space,
\be
\frac{\gamma}{(p-k)^2}\to\pi K_0^3\delta''(p-k),
\ee
with $K_0$ being the new constant with the dimension of mass. Due to the presence of the delta
function in $k$ in the interaction kernel, all integrals
can be considerably simplified. For example, the excess of the vacuum energy 
$\Delta\tilde{\cal E}_{\rm vac}[\theta]$, similar to (\ref{evac}), reads now:
\be
\Delta\tilde{\cal E}_{\rm vac}[\theta]=-\int\frac{dp}{2\pi}\left\{[p\sin\theta(p)-|p|]
-\frac{\pi K_0^3}{4}\left[\theta'(p)\right]^2\right\}.
\label{ev2}
\ee

The integral on the r.h.s. in (\ref{ev2}) is convergent and scales as
\be
\Delta\tilde{\cal E}_{\rm vac}[\theta]=\tilde{C}_1A^2+K_0^3\frac{\tilde{C}_2}{A}
\label{ev22}
\ee
under $\theta(p)\to\theta(p/A)$, so that qualitatively
the same conclusion holds true, that there exists only one phase of the theory with
spontaneously broken chiral symmetry and there is no unbroken phase. Similarly to
(\ref{EA3}), we can rewrite Eq.(\ref{ev22}) using the chiral condensate instead of
$A$:
\be
\Delta\tilde{\cal E}_{\rm vac}=\tilde{C}_1'\left[\frac12\left(\frac{\Sigma_{\phantom
0}}{\tilde{\Sigma}_0}\right)^2+\left|\frac{\tilde{\Sigma}_0}{\Sigma}\right|\right]
\label{EA32}
\ee
with the minimum given by $\Sigma=\tilde{\Sigma}_0$.

\begin{figure}[t]
\centerline{\epsfig{file=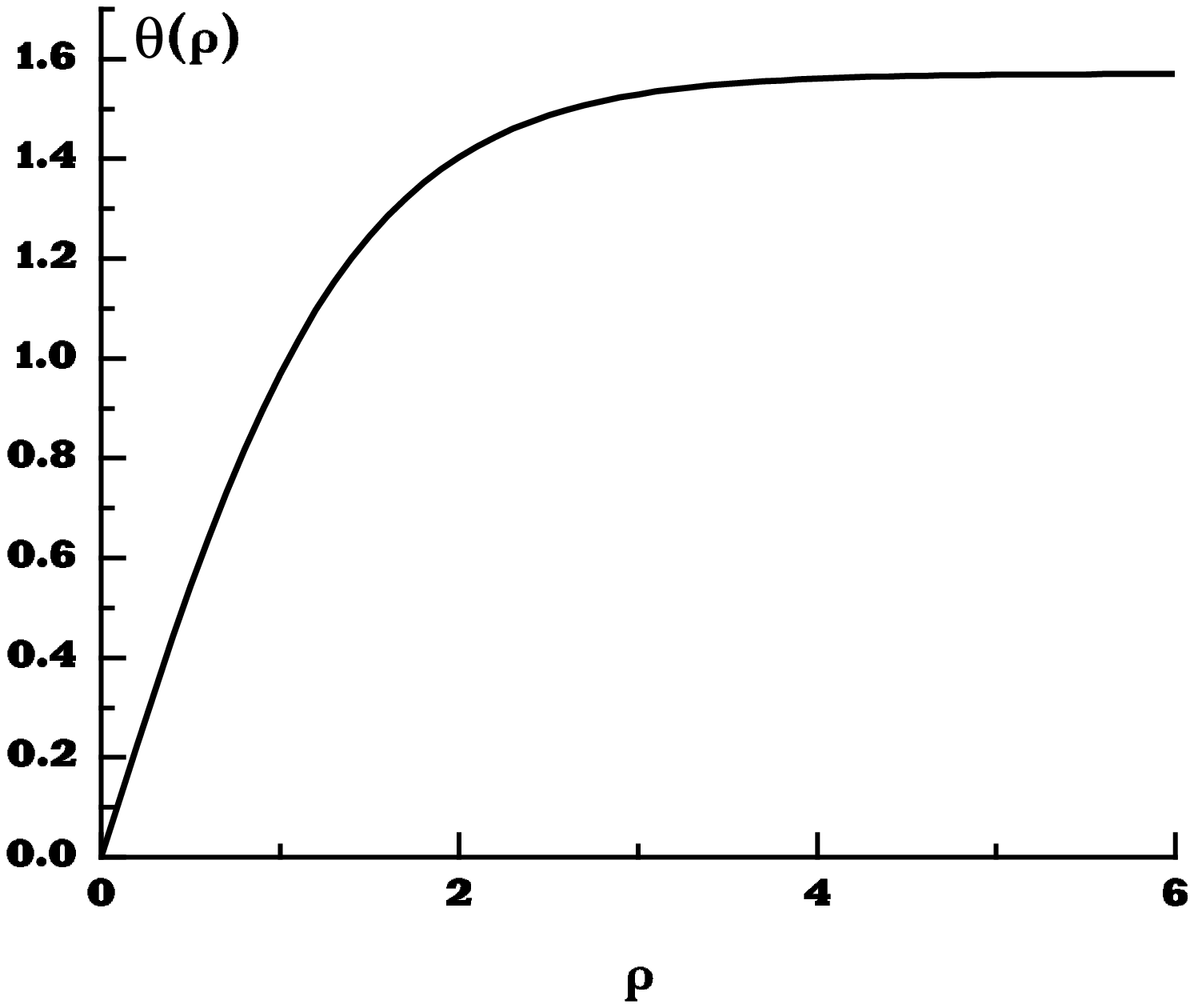,width=10.5cm}\hspace{-2.5cm}
            \epsfig{file=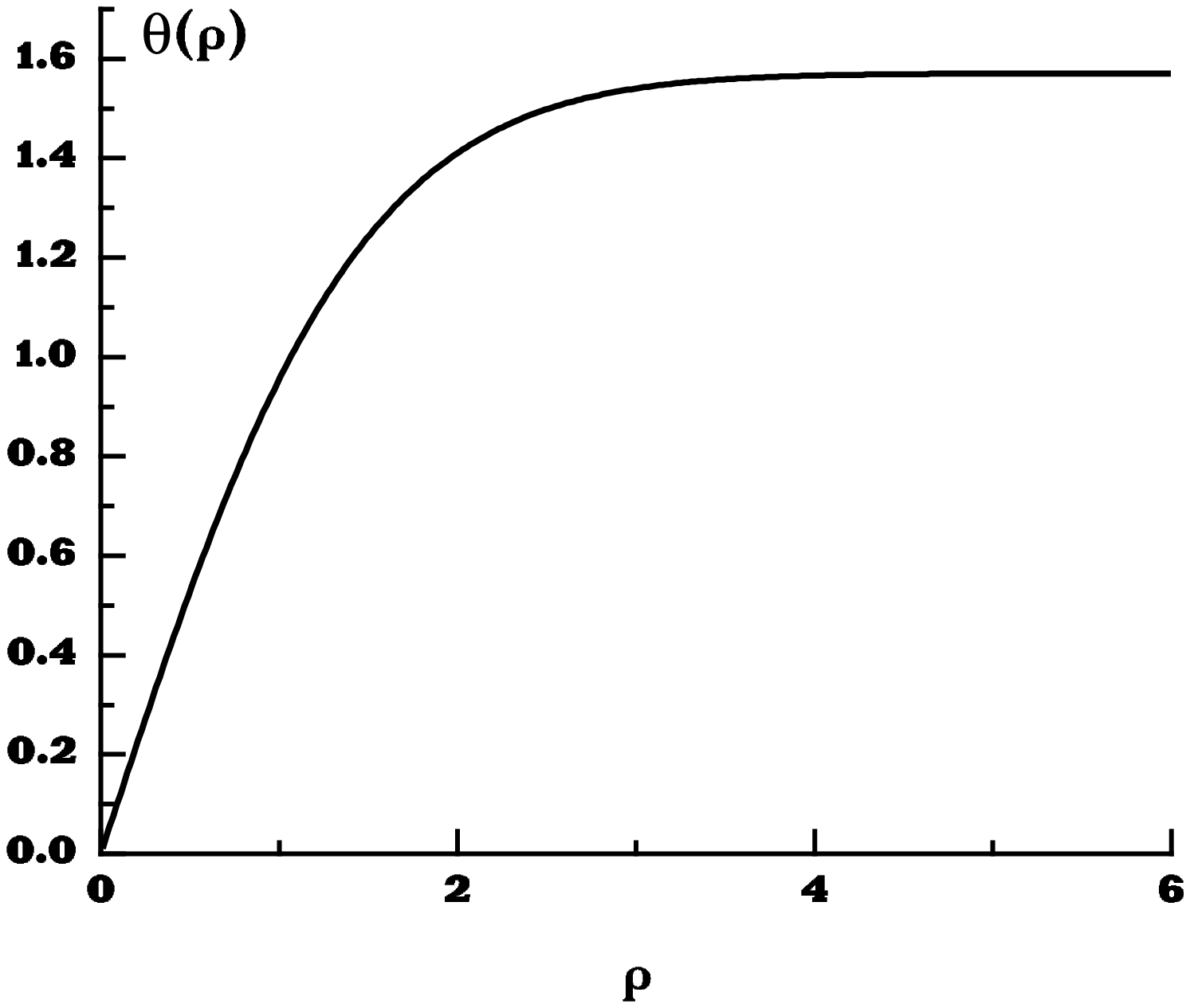,width=10.5cm}}
\caption{The ground-state solutions of the two-dimensional mass-gap equations (\ref{mg2})
which correspond to the linear (left plot) and the 
oscillator (right plot) potential, respectively. 
The momentum $p$ is given in units of $\sqrt{\gamma}$
for the left plot and in units of $K_0$ for the right one.}
\end{figure}

\subsection{Numerical solutions for the chiral angle}

In the previous subsections we found that there must be nontrivial functions $\theta(p)$
providing the minimum of the vacuum energies (\ref{evac}) and (\ref{ev2}). As was mentioned above, 
these functions are solutions to the mass-gap equations,
\be
\frac{\delta \Delta{\cal E}_{\rm vac}[\theta]}{\delta\theta(p)}=0,\quad
\frac{\delta \Delta\tilde{\cal E}_{\rm vac}[\theta]}{\delta\theta(p)}=0,
\ee
which take the following forms:
\be
p\cos\theta(p)=\frac{\gamma}{2}\int\frac{dk}{(p-k)^2}\sin[\theta(p)-\theta(k)],\quad
p\cos\theta(p)=-\frac{\pi K_0^3}{2}\theta''(p)
\label{mg2}
\ee
for the linear and harmonic-oscillator potential, respectively.

\begin{figure}[t]
\centerline{\epsfig{file=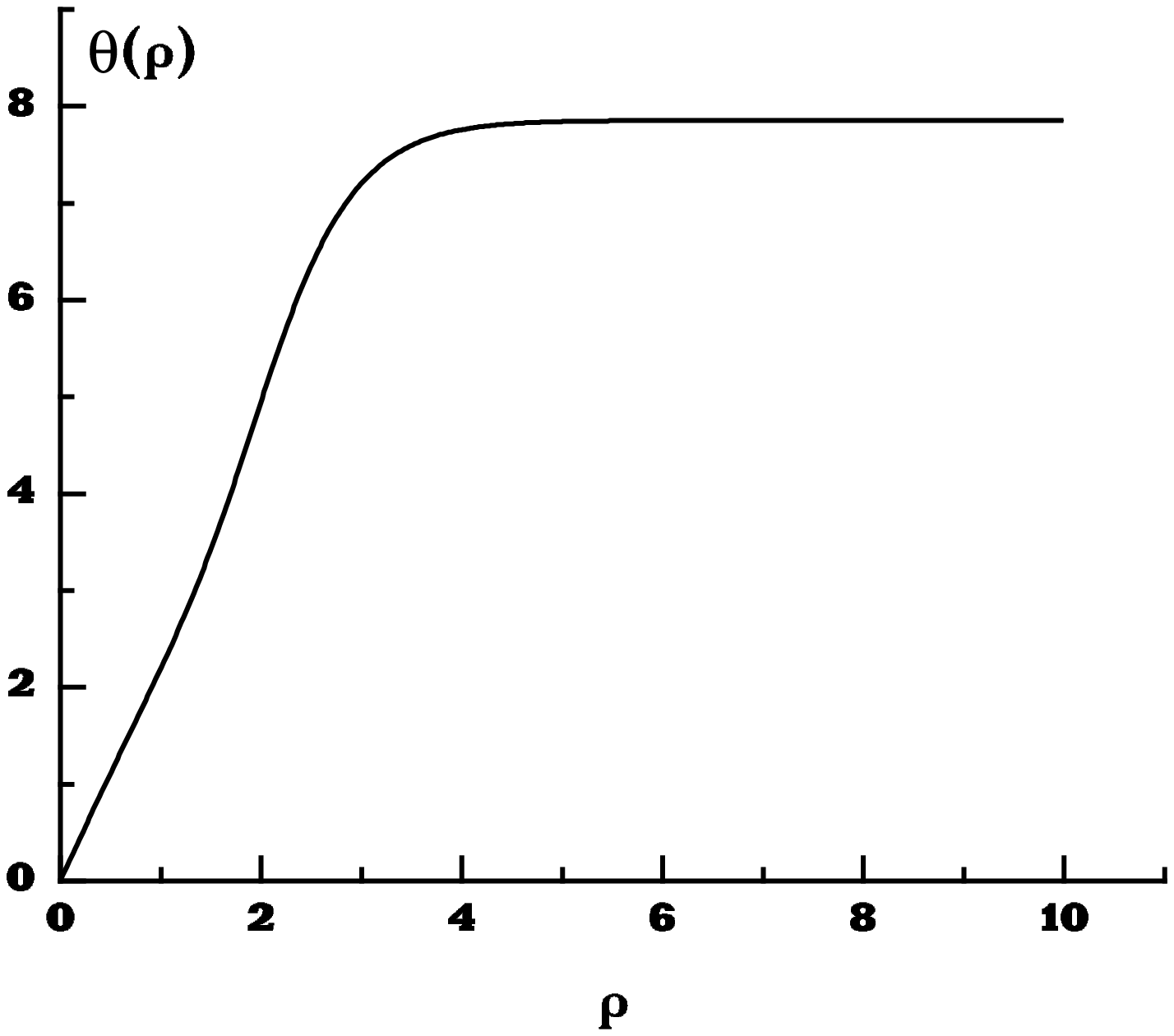,width=10.5cm}\hspace*{-2.5cm}\epsfig{file=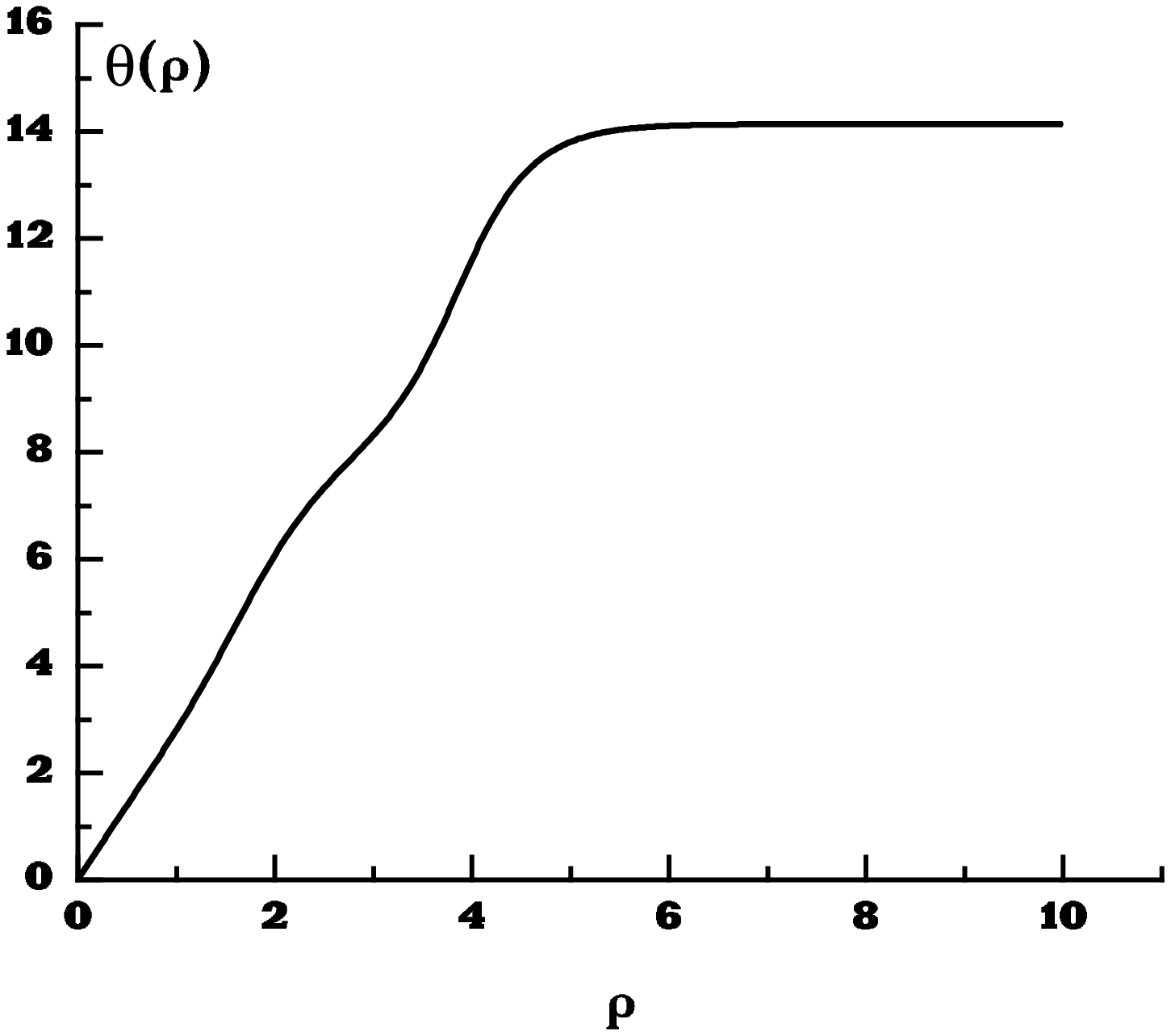,width=10.5cm}}
\caption{The unphysical excited solutions to the mass-gap equation (\ref{mg2}) (oscillator
confinement). The momentum $p$ is given in units of $K_0$.}
\end{figure}

Now we turn to numerical studies of these two equations. Note that the qualitative
analysis performed in the previous subsection 
cannot answer the question of how many solutions of the mass-gap equation exist. Indeed, a
second solution, if it exists, would just lead to different constants $C$'s ($\tilde{C}$'s)
and a different value of the chiral condensate $\Sigma_0$ ($\tilde{\Sigma}_0$), since 
this solution is
expected to have a different profile and it cannot be obtained from the ground-state
solution by dilatations. 

As the first step, we reproduce the solution of the first equation in (\ref{mg2}) 
found in \cite{Ming Li} and
construct a similar solution for the other (see Fig. 1). Notice that we
define the chiral angle $\theta(p)$ from $-\pi/2$ to $\pi/2$ with the boundary conditions
$\theta(p=0)=0$, $\theta(p\to\infty)\to\pi/2$ (as was mentioned before, 
$\theta(-p)=-\theta(p)$). It turns out that under such
conditions no excited solutions of the mass-gap equations (\ref{mg2}) exist.
Only if both requirements are relaxed, can one find the solutions depicted in Fig. 2, 
which start at 0 for
$p=0$ and tend to $\pi/2+2\pi n$, $n\ge 1$, for $p\to\infty$. From Fig. 1 one can easily see
that the solutions for the ground state in both types of the confining potential have a
very similar shape, so it would be natural to expect similar excited solutions to exist
for the linear potential as well. However, such
solutions lead to maximal condensation of the $\bar{q}q$ pairs not only at $p=0$, but for
some finite $p$'s, where $\cos\theta=1$, which is unphysical, so we disregard these
solutions. 

Thus we conclude that only the ground-state solution may exist for the mass-gap equation in
the 't~Hooft model, defining the only possible phase of the theory, with spontaneously
broken chiral symmetry and the chiral condensate
\be
\Sigma_0=-0.410N_C\sqrt{\gamma},
\label{condlin}
\ee
which coincides with the value found in a sequence of papers \cite{Ming Li,we1,Zhitnitsky}. 
A similar result for the theory with the oscillator-type interaction reads:
\be
\tilde{\Sigma}_0=-0.405N_CK_0,
\ee
which is very close to (\ref{condlin}) if the scales $\sqrt{\gamma}$ and $K_0$ are chosen
close to one another.

Notice that in the two-dimensional QCD the energy of the vacuum appears to be an apparently 
meaningless quantity by itself, although it leads to the meaningful mass-gap equation, as
discussed above. Indeed, the expression (\ref{evac}) contains a divergent
integral, so that the result depends on the cut-off. On the other hand, in view of the
fact that the physical solution of the mass-gap equations is unique, one cannot define a
difference of two energies, where the divergent term would cancel. 

Now, when the numerical methods are tuned and tested at the simple example of the
two-dimensional QCD, we turn to the main subject of the present paper --- namely, to
studies of the vacuum replicas in QCD$_4$.

\section{Four-dimensional QCD}

\subsection{Instantaneous interaction and forms of the potential. The mass-gap equation}

In this subsection we consider the case of QCD$_4$ and study the problem of existence of 
the vacuum
replicas for various types of gluonic correlators. In contrast to QCD$_2$, we put
the number of colours equal to three from the very beginning.

Following the standard procedure, we start from the QCD partition function and perform its
cluster expansion, so that, after integrating out the gluonic field, the theory contains
effective 4-quark, 6-quark, and so on vertices with the formfactors given by the
corresponding correlators of the gluonic fields,
\be
\langle\langle A_{\mu_1}^{a_1}(x_1)A_{\mu_2}^{a_2}(x_2)\rangle\rangle,\quad
\langle\langle
A_{\mu_1}^{a_1}(x_1)A_{\mu_2}^{a_2}(x_2)A_{\mu_3}^{a_3}(x_3)\rangle\rangle,\quad\ldots\;,
\ee
where the irreducible average, $\langle\langle\ldots\rangle\rangle$, is defined in the standard
way,
\be
\langle\langle 1\rangle\rangle=\langle 1\rangle,\quad 
\langle\langle 12\rangle\rangle=\langle 12\rangle-\langle 1\rangle\langle
2\rangle,\quad\ldots\;.
\ee

We fix the Fock-Schwinger gauge $(x-z_0)_{\mu}A_{\mu}^a(x)=0$ with an arbitrary fixed point
$z_0$ \cite{rg}, that allows us to connect the gluonic field with the field-strength
tensor, $F^a_{\mu\nu}(x)$,
\be
A^a_{\mu}(x)=\int_0^1 \alpha d\alpha (x-z_0)_{\nu} F^a_{\nu\mu}(\alpha(x-z_0)+z_0),
\ee
and, finally, to deal with the set of gauge-invariant cumulants \cite{Sim},
\be
\langle\langle Tr F_{\mu_1\nu_1}(x_1,z_0)F_{\mu_2\nu_2}(x_2,z_0)\rangle\rangle,\quad
\langle\langle
Tr F_{\mu_1\nu_1}(x_1,z_0)F_{\mu_2\nu_2}(x_2,z_0)F_{\mu_3\nu_3}(x_3,z_0)\rangle\rangle,\quad\ldots\;,
\label{Fs}
\ee
$$
F_{\mu\nu}^{a}(x,z_0)\equiv\Phi(z_0,x)F_{\mu\nu}^{a}(x)\Phi(x,z_0),
$$
with $\Phi(x,y)$ being the standard parallel transporter from the point $x$ to the point
$y$. Now we apply a set of approximations.

On one hand, the Gaussian dominance is used, which means that we keep only the 
low\-est-or\-der cor\-re\-la\-tor 
$\langle\langle FF\rangle\rangle$, whereas all others are considered suppressed \cite{Sim}. 
Such an approximation can be justified,
first, by the Casimir scaling, suggested long ago \cite{cs1} and recently confirmed by lattice
calculations \cite{cs2}. The Casimir scaling is exact in the Gaussian approximation.
On the other hand, the minimal-area law
asymptotic for an isolated Wilson loop can be saturated by the Gaussian correlator.
Finally, numerical calculations of hadronic observables within this method show 
good agreement with experimental and lattice data. The details of the method can be found,
{\it e.g.}, in the review \cite{Simrev} and references therein.

As far as the form of the nonperturbative part of the bilocal correlator is
concerned, we choose it in the form
$$
\langle\langle A_{\mu}^{a}(x)A_{\nu}^{b}(y)\rangle\rangle\hspace*{8cm}
$$
\be
=\int_0^1\alpha
d\alpha\int_0^1\beta d\beta (x-z_0)_{\lambda}(y-z_0)_{\sigma}
\langle\langle F^a_{\lambda\mu}(\alpha(x-z_0)+z_0)F^b_{\sigma\nu}(\beta(y-z_0)+z_0)
\rangle\rangle
\label{cr1}
\ee
$$
\hspace*{5cm}\to\delta^{ab}g_{\mu 0}g_{\nu 0}V_0(|\vec{x}-\vec{y}|)\delta(x_0-y_0),
$$
which implies that the gluonic correlation length $T_g$ (also playing the role of the QCD
string radius) is small enough, so that the string-like interaction can be approximated
by the instantaneous one \footnote{Lattice calculations \cite{DG} give the value 
$T_g\approx 0.2\div 0.3fm$
which is small enough at the hadronic scale.}. 
For the confining potential, $V_0(|\vec{x}|)$, 
we consider two cases,
 the quadratic and linear form,
\begin{eqnarray}
V_0(|\vec{x}|)=K_0^3|\vec{x}|^2,\label{v2}\\
V_0(|\vec{x}|)=\sigma_0 |\vec{x}|.\label{v1}
\end{eqnarray}
The Lorentz structure of confinement is taken to
be $\gamma_0\times\gamma_0$.
In the linear case we shall add to $V_0(|\vec{x}|)$ a Coulomb term and a constant, 
\be
V_1(|\vec{x}|)=-\frac{\alpha_s}{|\vec{x}|}+U,
\label{DV}
\ee
with the Lorentz structure being $\gamma_{\mu}\times\gamma_{\mu}$. Notice that due to the
colour structure of the interaction, being $\frac{\lambda^a}{2}\times\frac{\lambda^a}{2}$,
the constant term in the potential (\ref{DV}) yields a force between singlets which is
proportional to the Resonating Group Method normalization overlap of these singlets. This
is a simple example of the Wigner-Eckart theorem. The Coulomb potential is necessary for
this class of models to possess a heavy-quark limit compatible with the known
spectroscopy.

Thus our model is defined by the parameter set $\{\sigma_0,\;\alpha_s,\;U\}$. 
We shall choose the usual literature values for two of them, 
$\sigma_0=0.135GeV^2$, $\alpha_s=0.3$, and adjust $U$ to obtain a quark condensate 
$\langle\bar qq\rangle\simeq -(250MeV)^3$. 
Next we turn to the study of the vacuum structure following the same steps used 
in QCD$_2$. 

First, we introduce the dressed quark field \cite{Orsay,BR1},
\be
q_{\alpha}(t,\vec{x})=\sum_{\zeta=\uparrow,\downarrow}\int\frac{d^3p}{(2\pi)^3}e^{i\vec{p}\vec{x}}
[b_{\alpha \zeta}(\vec{p},t)u_\zeta(\vec{p})+d_{\alpha \zeta}^+(-\vec{p},t)v_\zeta(-\vec{p})],
\label{quark_field2}
\ee
with $\alpha$ and $\zeta$ being the quark colour and the projection of the spin,
respectively, and the amplitudes $u$ and $v$ defined as
\be
\begin{array}{l}
u(\vec{p})=\left[\sqrt{\frac{1+\sin\vp}{2}}+\sqrt{\frac{1-\sin\vp}{2}}\hat{\vec{p}}\cdot\vec{\alpha}\right]u_0(\vec{p}),\\
v(-\vec{p})=\left[\sqrt{\frac{1+\sin\vp}{2}}-\sqrt{\frac{1-\sin\vp}{2}}\hat{\vec{p}}\cdot\vec{\alpha}\right]v_0(-\vec{p}).
\end{array}
\ee 
Here the chiral angle $\vp$ is used instead of QCD$_2$ $\theta$. They are simply related by
\be
\vp(p)=\frac{\pi}{2}-\theta(p).
\ee  
It is convenient to define the new angle, $\vp$, from $-\pi/2$ to $\pi/2$ with the boundary
conditions $\vp(p=0)=\pi/2$ and $\vp(p\to\infty)\to 0$.

If the chiral angle $\vp(p)$ is not identically zero, then the true vacuum state is
populated by the quark-antiquark pairs with the $^3P_0$ coupling,
\be
|\tilde 0\rangle=\mathop{\prod}\limits_{cfp}dp\left(\frac{1+\cos2\phi(p)}{2}+
\frac{\sin2\phi(p)}{2}C^+_{cfp}+\frac{1-\cos{2\phi(p)}}{4}C^{+2}_{cfp}\right)|0\rangle
\equiv S_0|0\rangle,
\label{Sdef}
\ee
where indices $c$ and $f$ numerate the colour and the flavour, respectively;
the angle $\phi(k)$ is connected to the chiral one, $\vp(k)$, via a simple relation,
\be
\vp(p)=\arctan\frac{m}{p}+2\phi(p)\mathop{=}\limits_{m=0}2\phi(p),
\ee
and defines the above mentioned unitary operator $S_0$.
\be
C_{cfp}^+=(b^+_{\uparrow}(\vec{p}),b^+_{\downarrow}(\vec{p}))_{cf}\mathfrak{M}
\left(\begin{array}{c}d^+_{\uparrow}(\vec{p})\\
d^+_{\downarrow}(\vec{p})\end{array}\right)_{cf},
\ee
\be
\mathfrak{M}_{\sigma_1\sigma_2}=(-\sqrt{6})\sum_{\sigma,m}
\left(\begin{array}{ccc}1&1&0\\m&\sigma&0\end{array}\right)
\hat{\vec{p}}_{1m}\left(\begin{array}{ccc}1/2&1/2&1\\
\sigma_1&\sigma_2&\sigma\end{array}\right)
=\left(\begin{array}{cc}
-\hat{\vec{p}}_x+i\hat{\vec{p}}_y&\hat{\vec{p}}_z\\
\hat{\vec{p}}_z&\hat{\vec{p}}_x+i\hat{\vec{p}}_y\end{array}\right).
\ee

The operators annihilating the new vacuum, $|\tilde 0\rangle$, are
\be
\tilde{b}(\vec{p})=S_0b(\vec{p})S_0^+,\quad \tilde{d}(-\vec{p})=S_0d(-\vec{p})S_0^+.
\label{ops}
\ee
The interested reader can find the details of the formalism in papers \cite{BR1}.

The next step consists of arranging the normal ordering of the QCD$_4$ Hamiltonian,
\be
H=\int d^3 x[H_0(\vec{x})+H_I^{(1)}(\vec{x})+H_I^{(2)}(\vec{x})],
\ee
\be
H_0(\vec{x})=q^+(\vec{x})\left(-i\vec{\alpha}\cdot\vec{\bigtriangledown}\right)q(\vec{x}),
\ee
\be
H_I^{(1)}(\vec{x})=\frac12\int d^3
y\;q^+(\vec{x})\frac{\lambda^a}{2}q(\vec{x})\;\biggl[V_0(|\vec{x}-\vec{y}|)+V_1(|\vec{x}-\vec{y}|)\biggr]\;q^+(\vec{y})\frac{\lambda^a}{2}q(\vec{y}),
\ee
\be
H_I^{(2)}(\vec{x})=-\frac12\int d^3
y\;q^+(\vec{x})\vec{\alpha}\frac{\lambda^a}{2}q(\vec{x})\;V_1(|\vec{x}-\vec{y}|)
\;q^+(\vec{y})\vec{\alpha}\frac{\lambda^a}{2}q(\vec{y}).
\ee

In the new basis, we obtain an expression similar to (\ref{Hh}) and we again concentrate
on the minimization of the vacuum-energy density --- namely, on the mass-gap equation,
which takes the form:
\be
A(p)\cos\vp(p)-B(p)\sin\vp(p)=0,
\label{mge}
\ee
$$
A(p)=\frac12C_F\int
\frac{d^3k}{(2\pi)^3}\;\biggl[V_0(\vec{p}-\vec{k})+4V_1(\vec{p}-\vec{k})\biggr]\sin\vp(k),
$$
$$
B(p)=p+\frac12C_F\int \frac{d^3k}{(2\pi)^3}\;
(\hat{\vec{p}}\cdot\hat{\vec{k}})\;\biggl[V_0(\vec{p}-\vec{k})+2V_1(\vec{p}-\vec{k})\biggr]\cos\vp(k),
$$
where $C_F=\frac43$ is the Casimir operator in the fundamental representation.

Thus, equipped with the general form of the mass-gap equation, we are in the position to
perform the numerical search for the excited vacua for the oscillator and linear
potentials.

\subsection{Solutions of the mass-gap equation for the oscillator-type interaction}

In this subsection we solve numerically the mass-gap equation (\ref{mge}) for the
oscillator-type confining potential (\ref{v2}) and $V_1$ put to zero. 
Similarly to the two-dimensional case, the Fourier transform of the potential (\ref{v2})
is the second derivative of the delta function,
\be
V_0(\vec{p})=-(2\pi)^3K_0^3\frac{\partial^2}{\partial \vec{p}^2}\delta^{(3)}(\vec{p}),
\ee
which leads to the mass-gap equation in the differential form,
\be
\vp''=-\frac{2}{p}\vp'+\frac{2p}{C_FK_0^3}\sin\vp-\frac{1}{p^2}\sin2\vp,
\label{diffmge}
\ee
and we solve it numerically using the Runge-Kutta method. In Fig. 3 we show the ground-state and
the first excited solution of Eq.(\ref{diffmge}).

Thus we conclude for the existence of  at least one replica for the case of quadratic confining
potential. Moreover, using the same numerical technique, one can build the third, the
fourth, and so on excited solutions, as was found in \cite{ORS2}. 
In fact, the harmonic oscillator supports an infinite tower of such replicas.

\begin{figure}[t]
\centerline{\epsfig{file=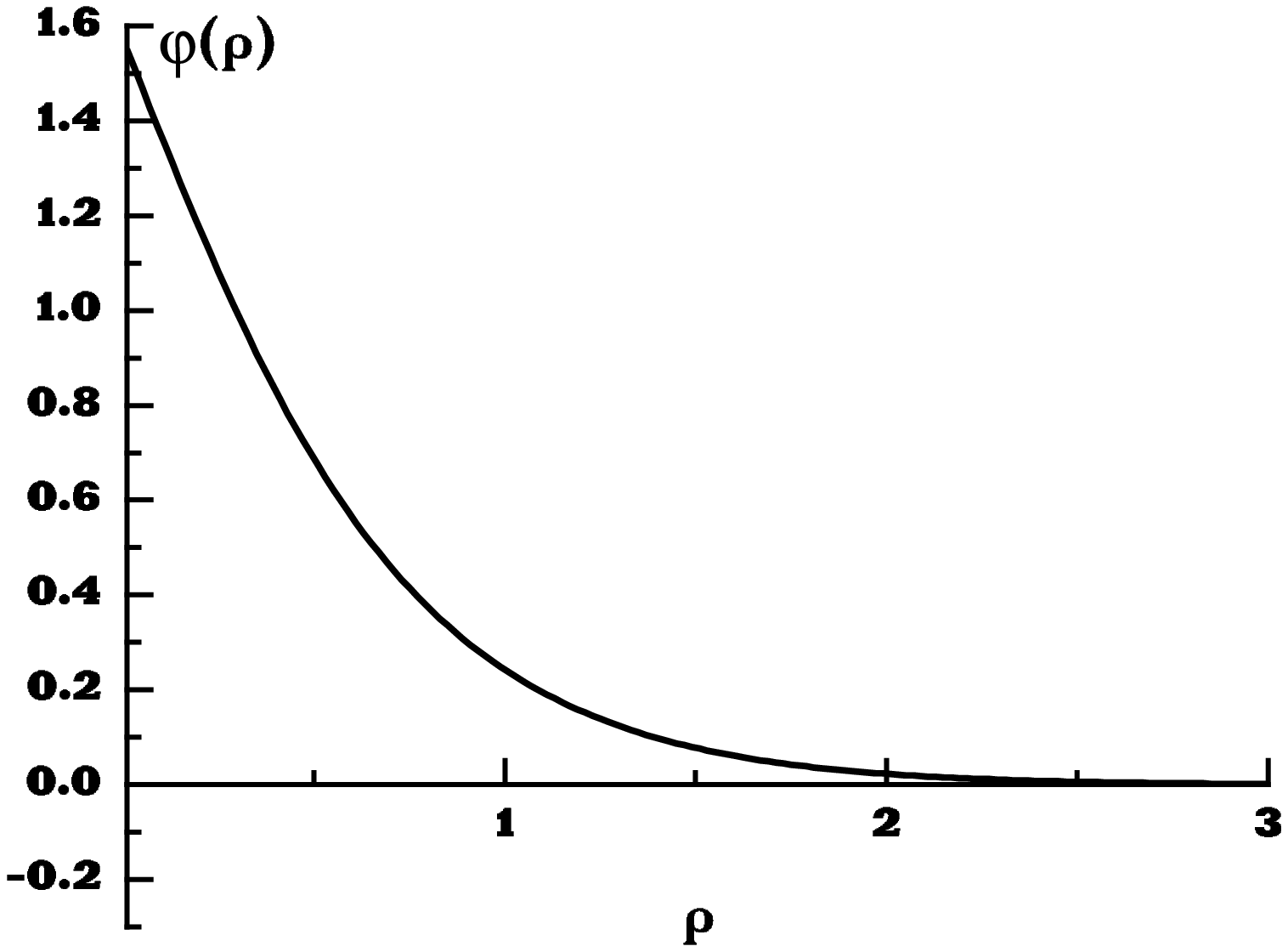,width=10.5cm}\hspace{-2.5cm}
            \epsfig{file=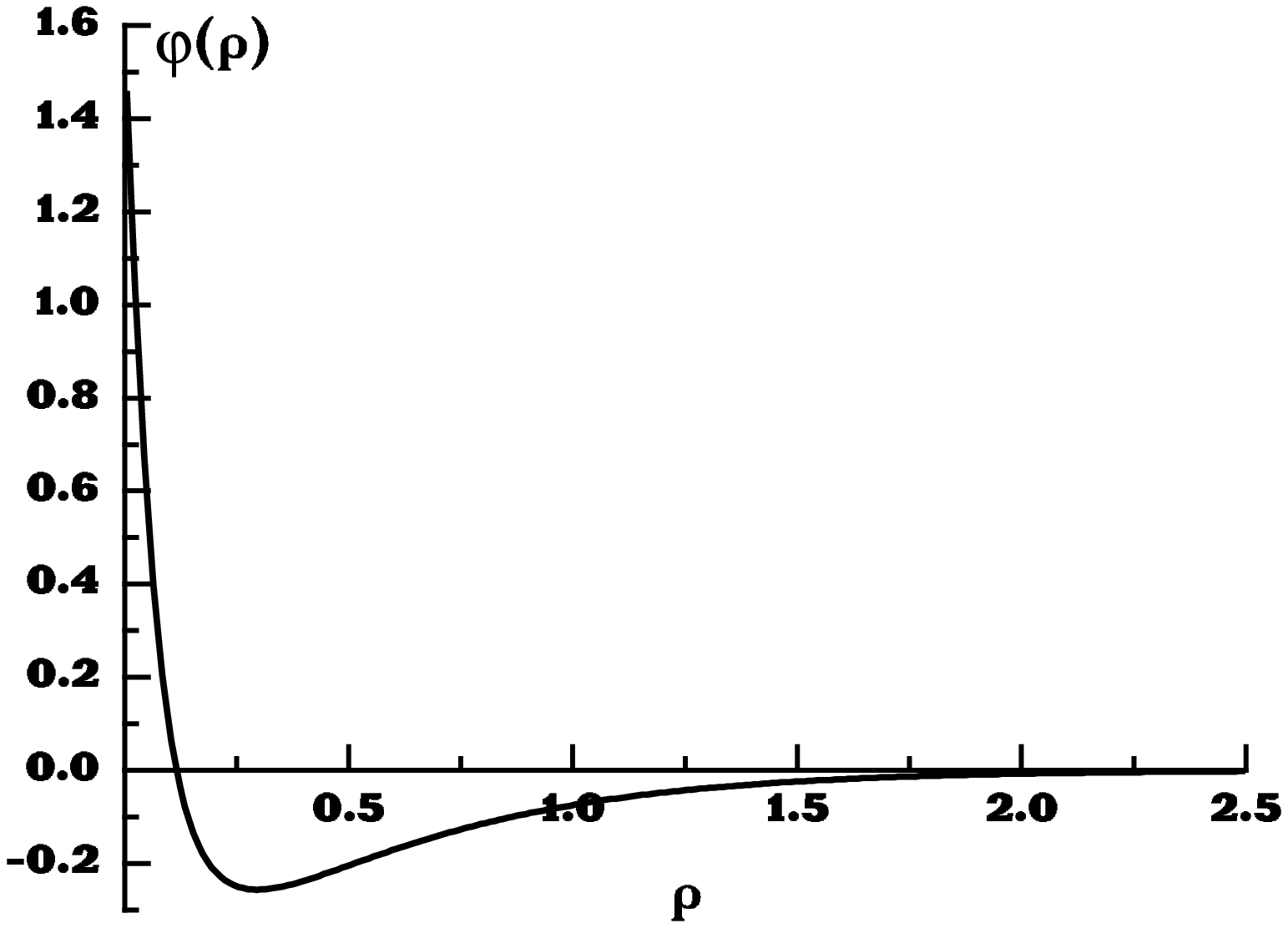,width=10.5cm}}
\caption{The first two solutions of Eq.(\ref{diffmge}) for the chiral angle $\vp(p)$
found with the Runge-Kutta method. The momentum $p$ is given in units of $K_0$.}
\end{figure}

\subsection{Vacuum replica for the linear potential}

Finally, we turn to the main goal of the present paper --- namely, whether the linear confining potential 
supports the existence of replicas.

In contrast to the case of the oscillator potential, the mass-gap equation for the linear
confinement, given by Eq.(\ref{v1}), is integral,
\be
p\sin\vp(p)=\frac{C_F\sigma_0}{2\pi^2}\int\frac{d^3k}{(\vec{p}-\vec{k})^4}\left[
\sin\vp(k)\cos\vp(p)-\hat{\vec{p}}\cdot\hat{\vec{k}}\sin\vp(p)\cos\vp(k)\right],
\label{V0mg}
\ee
or, equivalently,
\begin{eqnarray}
\label{lmge}
p\sin\vp(p)&=&\frac{\sigma}{4\pi}\int_0^{\infty}dk\;k^2\left[\frac{4}{(p^2-k^2)^2}
\sin\vp(k)\cos\vp(p)\right.\\
&&\left.-\left(\frac{2k(k^2+p^2)}{p(p^2-k^2)^2}-
\frac{1}{p^2}\ln\left|\frac{p+k}{p-k}\right|\right)\cos\vp(k)\sin\vp(p)\right]\nonumber,
\end{eqnarray}
where $\sigma=C_F\sigma_0$ is the string
tension in the fundamental representation, and the angular variables are integrated out. 
One can check that the integral on the r.h.s. of (\ref{lmge}) is convergent.  Eq.(\ref{lmge})
possesses only one nontrivial eigenstate: the ground-state vacuum. Everything happens as if 
the confining linearly rising potential just fails to be \lq\lq binding" enough to
produce an excited vacuum replica. Besides, the chiral condensate appears strongly
underestimated in case of the pure linear confining potential \cite{linear}. 
Then we supply it with the extra interaction, given by
(\ref{DV}), which reads, in momentum space,
\be
V_1(\vec{p})=-\frac{4\pi\alpha_s}{\vec{p}^2}+(2\pi)^3U\delta^{(3)}(\vec{p}),
\label{V11}
\ee 
and modify the mass-gap equation accordingly,
\begin{eqnarray}
\label{lmge2}
p\sin\vp(p)&=&\frac{\sigma}{4\pi}\int_0^{\infty}dk\;k^2\left[\frac{4}{(p^2-k^2)^2}
\sin\vp(k)\cos\vp(p)\right.\\
&&\left.-\left(\frac{2(k^2+p^2)}{pk(p^2-k^2)^2}-
\frac{1}{p^2k^2}\ln\left|\frac{p+k}{p-k}\right|\right)\cos\vp(k)\sin\vp(p)\right]\nonumber\\
&&-\frac{C_F\alpha_s}{4\pi}\int_0^{\infty}dk\;k^2\left[\frac{4}{pk}\ln\left|\frac{p+k}{p-k}\right|
\sin\vp(k)\cos\vp(p)\right.\nonumber\\
&&\left.+\left(\frac{p^2+k^2}{p^2k^2}\ln\left|\frac{p+k}{p-k}\right|-\frac{2}{pk}\right)\cos\vp(k)\sin\vp(p)\right]\nonumber\\
&&+\frac14C_FU\sin2\vp(p)\nonumber.
\end{eqnarray}

In contrast to Eq.(\ref{lmge}), the mass-gap equation (\ref{lmge2}) is not
divergence-free because of the Coulomb part, proportional to $\alpha_s$.
If we considered the case of a nonzero quark mass $m$, then the given
divergence would be only logarithmic, however, for Eq.(\ref{lmge2}) 
we have $m=0$ and the divergence becomes linear. We regularize it by using a modified 
potential $V_1$,
\be
V_1(\vec{p})=4\pi\alpha_s\left[\frac{1}{\vec{p}^2+\Lambda^2}-\frac{1}{\vec{p}^2}\right]+(2\pi)^3U\delta^{(3)}(\vec{p}),
\label{V1mod}
\ee 
with a cut-off $\Lambda$. Notice that the mass-gap equation 
for the single Coulomb potential does not contain any scale 
(it would be given by the quark mass $m$ beyond the chiral limit), which
appears only after the regularization and is given by $\Lambda$. Then all 
quantities with the
dimension of mass are directly proportional to $\Lambda$.

Now let us assume, for a moment, the general case of an arbitrary nonvanishing current 
quark mass. Then the quarks acquire an effective mass
$m_{\rm eff}$, which can be written as a sum of two components, 
\be
m_{\rm eff}=m+\Sigma(m,\Lambda),
\label{meff1}
\ee
where $m$ and $\Sigma(m,\Lambda)$ are the current quark mass and the quark
selfenergy due to the contribution of the chiral condensate, respectively. The latter
is obviously $\Lambda$-dependent. To evaluate $m_{\rm eff}$ one can use the relations
\be
\sin\vp(p)=\frac{m(p)}{\sqrt{p^2+m^2(p)}},\quad \cos\vp(p)=\frac{p}{\sqrt{p^2+m^2(p)}},
\label{efm}
\ee
similar to the free-case relations, but with $\vp(p)$ being the solution to the mass-gap
equation. Now the effective mass, $m_{\rm eff}$, can be associated with $m(p\sim 0)$.
Let us consider the heavy-quark limit, $m\to\infty$, first. A large quark current 
mass is known to
destroy the chiral condensate. As a result, the self-energy part of the relation
(\ref{meff1}) must vanish, so that the effective mass is basically given by the 
current mass,
\be
m_{\rm eff}\mathop{\simeq}\limits_{m\to\infty}m.
\ee
In particular, 
$\Sigma(m,\Lambda)/m\mathop{\to}\limits_{m\to\infty}0$. That is, for large $m$, we have 
to go
up in $\Lambda$ for the system to become more and more Coulombic, whereas the chiral
condensate goes to zero. In other words the potential of Eq.(\ref{V1mod}) naturally 
possesses the Coulomb limit for very heavy current quark masses.

In the opposite limit, $m=0$, the effective mass appears entirely due to the chiral 
symmetry breaking,
\be
m_{\rm eff}=\Sigma(0,\Lambda)=\Lambda\times Const,
\label{meff2}
\ee 
where the last equality holds true provided we choose 
$\Lambda\sim\sqrt{\sigma_0}$, the only scale of
the problem. Thus in {\em either} case $m_{\rm eff}\gtrsim\Lambda$, and it is remarkable
that such a potential should equally address both heavy- and light-quark sectors.

Therefore we adopt the following 
strategy. We choose the standard values of the string tension, $\sigma_0=0.135GeV^2$, and the 
strong coupling constant, $\alpha_s=0.3$. As discussed above, we put $\Lambda$ to be of order
of $\sqrt{\sigma_0}$ and solve the resulting mass-gap equation
numerically for the ground state, fitting the only free parameter, $U$, in such a way, 
as to have the chiral condensate,
\be
\langle \bar qq\rangle=-\frac{3}{\pi^2}\int_0^{\infty}dp\;p^2\sin\vp(p),
\label{qqq}
\ee
to be around its standard value, about 
$-(250MeV)^3$. With the solution found for the
chiral angle $\vp(p)$, we evaluate the effective quark mass, $m_{\rm eff}$, and check if it
has the value of order $200\div 300MeV$, that is, 
the value of the same order of magnitude as the regulator $\Lambda$, being also
close to the value usually adopted for
the constituent quark mass. For selfconsistency we also need to ensure that
the fitting parameter, $U$, also appears to be of order of the interaction scale and,
hence, does not bring a new scale into the problem. In Table~1, besides the known values
of $\sigma_0$ and $\alpha_s$, we give the best set of parameters $U$ and $\Lambda$
consistent with the standard value of the chiral condensate. 

\begin{table}[t]
\caption{Parameters of the model fixed from the fit for the ground vacuum state.}
\begin{ruledtabular}
\begin{tabular}{cccccc}
$\sigma_0$, $GeV^2$&$\alpha_s$&$U$, $MeV$&$\Lambda$, MeV&$m_{\rm eff}$, $MeV$&
$\langle \bar qq\rangle$, $MeV^3$\\
0.135&0.3&220&250&230&-$(250)^3$\\
\end{tabular}
\end{ruledtabular}
\end{table}

Now, when all parameters of the model are
totally fixed, we perform our search for the excited vacuum replica which, indeed, 
appears to exist, and we give its profile, together with the ground-state one, in Fig. 4.
We tried many other
combinations of $U$ and $\Lambda$ and, whatever values we have used, 
we always found this replica to persist. 

\begin{figure}[t]
\centerline{\epsfig{file=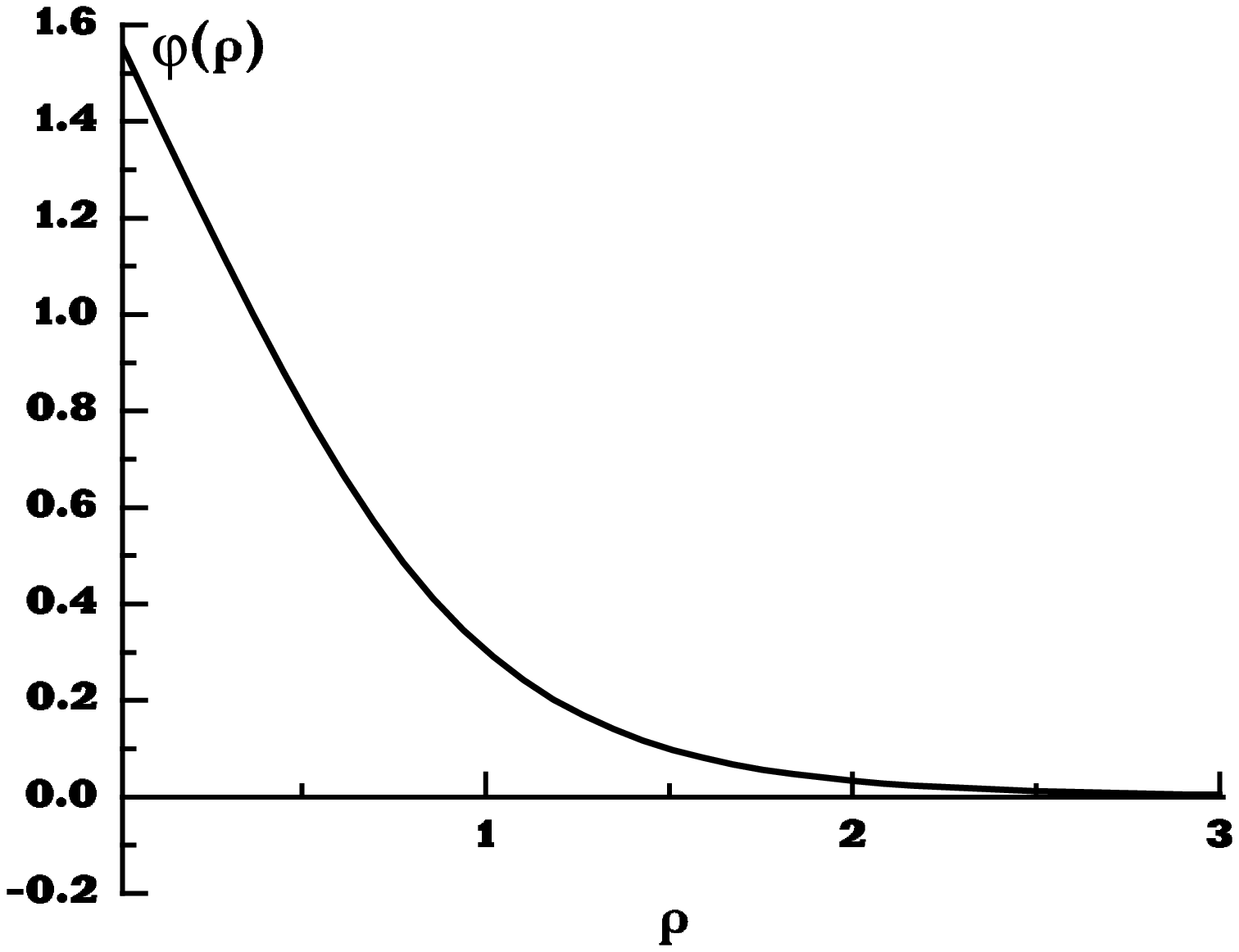,width=10.5cm}\hspace*{-2.5cm}
            \epsfig{file=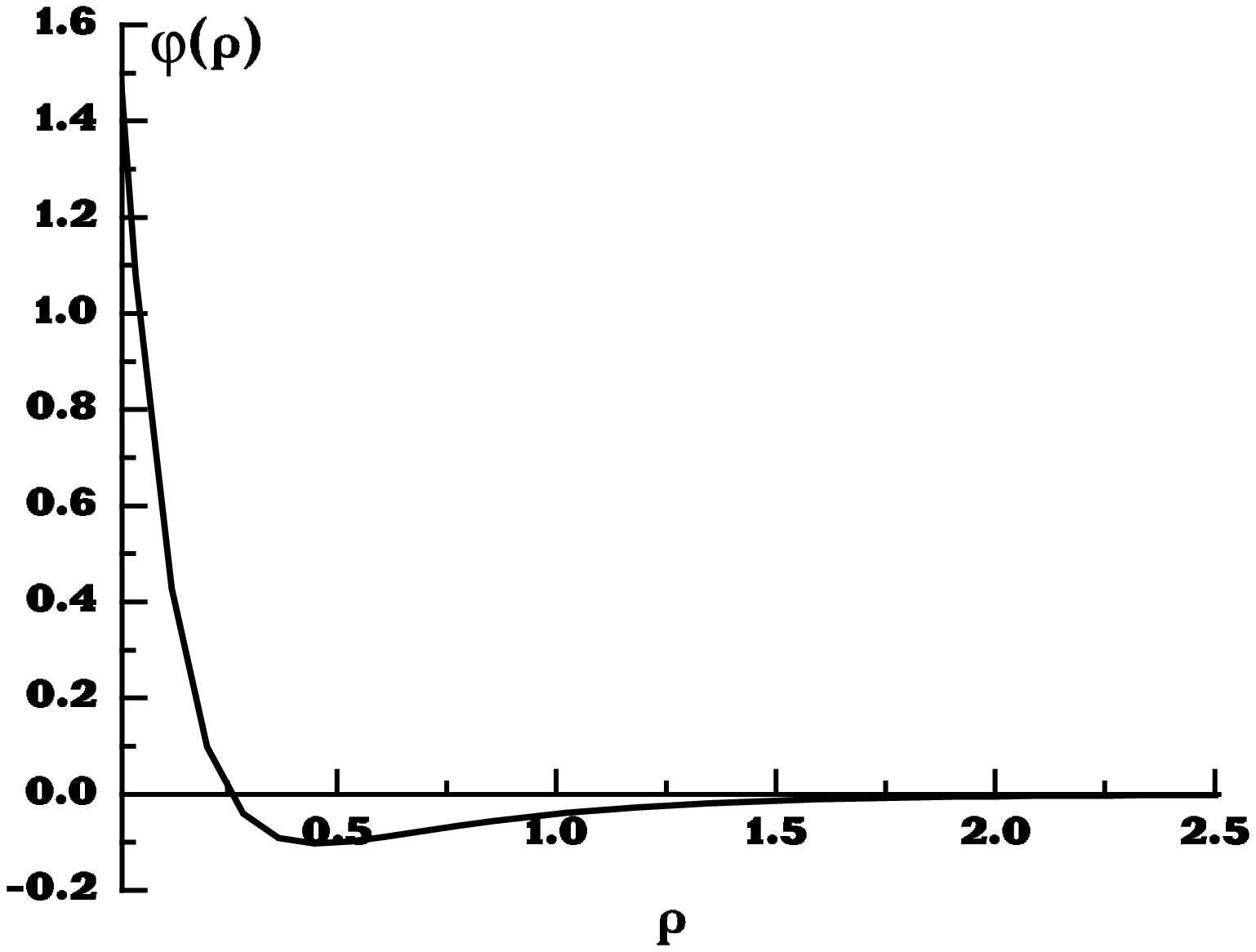,width=10.5cm}}
\caption{The ground-state (left plot) and the first excited (right plot) solution of 
Eq.(\ref{lmge2}) for the set of parameters given in Table~I. The momentum $p$ is given in
units of $\sqrt{\sigma_0}$.}
\end{figure}

The chiral condensate for the excited vacuum state,
\be
\langle\widetilde{\bar qq}\rangle=(126MeV)^3,
\label{conpr}
\ee
is smaller in the absolute value and has a \lq\lq wrong" sign compared to the
ground-state one, given in Table~I.
For massless quarks this sign can be easily reversed by the
change $\vp\to-\vp$ everywhere, which also satisfies the mass-gap equation and does not affect
the vacuum energy. However, nonvanishing quark mass breaks this symmetry, since an extra
term, $m\cos\vp(p)$, appears in the mass-gap equation (\ref{V0mg}). It follows immediately from the
Gell-Mann-Oakes-Renner relation that the sign of the condensate given in (\ref{conpr})
leads to the pion with an imaginary mass, that is, to the tachyon. However, the 
latter does not
lead to a disaster since the excited vacuum state is, indeed, unstable and
it tends to decay to the ground state.

Now we calculate the density of the vacuum energy defined as
\be
\Delta{\cal E}_{\rm vac}=-\frac{g}{2}\int\frac{d^3p}{(2\pi)^3}
\biggl(A(p)\sin\vp(p)+[B(p)+p]\cos\vp(p)\biggr)-{\cal E}_{\rm vac}^{\rm trivial},
\ee
where $A(p)$ and $B(p)$ are given in (\ref{mge}) and the degeneracy factor $g$ counts the
number of independent quark degrees of freedom,
\be
g=(2s+1)N_CN_f,
\ee
with $s=\frac12$ being the quark spin; the number of colours, $N_C$, is put to three, and
the number of light flavours, $N_f$, is two. Thus we find that $g=12$. The trivial 
solution of the mass-gap equation is $\vp_{\rm trivial}(p)\equiv 0$ (in contrast to the
two-dimensional case, this solution does not lead to difficulties with the vacuum energy).
Then for the two nontrivial solutions, depicted in Fig. 4, one has
\be
\Delta{\cal E}_{\rm vac}^{\rm ground}=-(128MeV)^4
\label{vE1}
\ee
and
\be
\Delta{\cal E}_{\rm vac}^{\rm excited}=-(36MeV)^4.
\label{vE2}
\ee
The negative sign ensures that they both are more energetically preferable than the trivial
vacuum.

It is instructive to rewrite Eqs.(\ref{vE1}), (\ref{vE2}) in the form of the vacuum
energy per Fermi cubed:
\be
\Delta{\cal E}_{\rm vac}^{\rm ground}=-34.7MeV/fm^3,
\label{vE12}
\ee
\be
\Delta{\cal E}_{\rm vac}^{\rm excited}=-0.2MeV/fm^3.
\label{vE22}
\ee

Now, if the vacuum is locally excited, then the extra energy stored in the volume 
${\cal V}_0=1fm^3$ is 
\be
\Delta E_0=\left(\Delta{\cal E}_{\rm vac}^{\rm excited}-
\Delta{\cal E}_{\rm vac}^{\rm ground}\right){\cal V}_0\approx 34.5MeV.
\label{trista}
\ee

We have checked the result (\ref{trista}) against slight variations of the parameters of
the model and found it to be very stable, changing only by several per cent.

\section{Physical processes in presence of the excited vacuum replica}

In this section we show how the excited vacuum could be \lq\lq seen" in hadronic
processes. 

As follows from Eq.(\ref{trista}), the energy stored in the bubble with 
the excited vacuum inside is proportional to its volume. At the moment
we can only give hand-waving qualitative arguments
concerning the most probable value of this volume. Indeed, the energy scale of the 
interaction we have in QCD is about $300MeV$ that corresponds to the distances
about $0.5fm$ which can be expanded up to the Compton length of the pion, 
being about $1.5fm$ or so --- the standard hadronic scale.
With the bubble of such a radius one can achieve the energy around $300MeV$, 
which is sufficient to produce an S-wave pair of pions when the excited vacuum decays
back to the ground-state one. This value looks quite naturally since all parameters of the
model, like the interaction scale, $\sqrt{\sigma_0}$, and the SBCS one,
given by the chiral condensate or by $m_{\rm eff}$ (see Table I)
have the same order of magnitude. As was discussed above, all three scales should always
appear selfconsistently, regardless of the details of the model used for
numerical calculations. In the present paper we give just an example of such a model, 
which we find rather realistic. 

It is instructive to see how the excited vacuum enters the amplitudes of the hadronic
processes. Usually one is to evaluate the matrix element of the form
\be
\langle {\tilde 0},hadrons|{\cal O}|{\tilde 0},hadrons' \rangle
\label{meop}
\ee
with ${\cal O}$ being a field operator. Note that this is the true vacuum of the theory, 
$|\tilde 0 \rangle$, to enter the matrix element, and we put it explicitly in 
(\ref{meop}). If the excited vacuum comes into the game, then it can appear either in the 
{\it bra} or in the {\it ket} vector, or in both. We use the notation 
$|\tilde 1 \rangle$ for it. Let us illustrate this statement by the
correlator of two quark electromagnetic currents,
\be
\langle {\tilde 0}|TJ_{\mu}(x)J_{\nu}(y)|{\tilde 1}\rangle
=\langle {\tilde
0}|Te^{i\hat{P}x}J_{\mu}(0)e^{-i\hat{P}x}e^{i\hat{P}y}J_{\nu}(0)e^{-i\hat{P}y}|{\tilde
1}\rangle,
\label{cr}
\ee
where $\hat{P}$ is the operator of the total momentum. Now, disentangling the $T$ product
and making the external operators $e^{\pm i\hat{P}x}$ and $e^{\pm i\hat{P}y}$ act on the
vacuum states to the left or to the right, one can arrive at the expression
\be
\langle {\tilde 0}|TJ_{\mu}(x)J_{\nu}(y)|{\tilde 1}\rangle=
\left\{
\begin{array}{ll}
e^{-i\Delta Py}\langle {\tilde 0}|J_{\mu}(x-y)J_{\nu}(0)|{\tilde 1}\rangle,&x_0>y_0\\
e^{-i\Delta Px}\langle {\tilde 0}|J_{\mu}(y-x)J_{\nu}(0)|{\tilde 1}\rangle,&y_0>x_0,
\end{array}
\right.
\label{correl}
\ee
where $\Delta P$ is the difference of the four-momenta of the two vacua, which, actually,
has only the zeroth component --- the difference of the vacuum energies, $\Delta E$. If
the Fourier transform is now applied to (\ref{correl}),
\be
{\cal P}_{\mu\nu}(p,q)=\int d^3x d^3y e^{ipx}e^{-iqy}\langle {\tilde
0}|TJ_{\mu}(x)J_{\nu}(y)|{\tilde 1}\rangle,
\ee
then 
\be
{\cal P}_{\mu\nu}(p,q)\sim\delta(p_0-q_0-\Delta E)\delta^{(3)}(\vec{p}-\vec{q}),
\ee
that is, one has an effective source pumping the energy into the system. This energy can
be spent to produce photons on mass shell each with an energy 
$\omega=\Delta E/2\simeq 10MeV$ for the radius of the excited-vacuum bubble being of 
order of the interaction radius, that is, of order about $0.5fm$. 

Notice that if the BCS vacuum structure is ignored and the trivial
vacuum, $|0\rangle$, is used instead of the $|{\tilde 0}\rangle$ and $|{\tilde 1}\rangle$ in
(\ref{cr}) (this corresponds to the chiral angles $\vp$ and $\tilde{\vp}$ identically
put to zero), then the latter expression coincides with the standard perturbative-vacuum
polarization operator supplied by the external energy source. In order to perform
calculations taking into account the nontrivial functions $\vp$ and $\tilde{\vp}$, one
should express all quantities in terms of the quark creation and annihilation operators 
inherent to one and the same vacuum, and to apply the standard Wick contractions then.
For example, relating everything to the trivial-vacuum operators, one can find:
\be
|{\tilde 0}\rangle=S_0|0\rangle,\quad |{\tilde 1}\rangle=S_1|0\rangle 
\ee
with operator $S_0$ defined in (\ref{Sdef}) and a similar definition of $S_1$
with the obvious change $\vp\to\tilde{\vp}$. Then the correlator (\ref{correl}) reads:
\be
\langle {\tilde 0}|TJ_{\mu}(x)J_{\nu}(y)|{\tilde 1}\rangle=
\frac{\langle 0|S_0^+\;TJ_{\mu}(x)J_{\nu}(y)\;S_1|0\rangle}{\langle 0|S_0^+S_1|0\rangle},
\label{Cornew}
\ee
where we used the correct normalization which removes the disconnected diagrams.
Note that in (\ref{Cornew}) the currents are also expected to be expressed in terms
of the trivial-vacuum quark creation/annihilation operators, so that, starting from
the quark field (\ref{quark_field2}), one must use transformations (\ref{ops}) (and
similar ones for the excited vacuum replica) to have everything selfconsistently.

An important comment is in order here. In the above analysis we did not consider the
mechanism of the vacuum excitation from the ground state to the replica. 
Indeed, to excite the vacuum inside of some region
in space, one should have a trigger --- a localized operator with the vacuum quantum
numbers.
Such an operator could be constructed out of the background gluonic fields (see 
Eq.(\ref{Fs}) above) -- the only
bricks at our disposal. If so, the outgoing photons or pions,
produced as a result of the excited vacuum decay, have to carry information about the
vacuum-gluonic-field correlations and, if measured experimentally, offer a possibility of
the direct probes of the QCD vacuum structure!

\section{Conclusions and outlook}

In the present paper we study the properties of the vacuum states in the potential
models for QCD. We find that there is only one vacuum in the two-dimensional case, and it
provides spontaneous breaking of the chiral symmetry. Lack of extra dimensions prevents
this simple model from having vacuum replicas. On the contrary, the
four-dimensional theory reveals a richer vacuum structure, not only containing the trivial
solution with unbroken chiral symmetry, but also possessing excited vacuum states, which
we find numerically for the oscillator and linear confining force. In the latter case
the Coulomb and the constant potential were also added to have a better correspondence
with real QCD. Thus we draw the following picture. Starting from the theory with bare quarks and
performing the Bogoliubov-Valatin transformation to \lq\lq dress" them, we find several
different types of this \lq\lq dressing", which correspond to different vacuum
states and differently broken chiral symmetry. The mostly broken phase, that is, the
one with the maximal (in the absolute value) chiral condensate and the solution for the chiral 
(Bogoliubov) angle without knots, possesses the minimal vacuum energy and, thus, defines the true vacuum of
the theory. Besides, there are other solutions of the mass-gap
equation with one, two, and more knots, defining other phases of the theory, which are less
energetically preferable than the ground state, but still possess lower vacuum energy than
the trivial vacuum. All these excited solutions lead to spontaneous breaking of the chiral
symmetry, though the chiral condensate decreases (in the absolute value) for each next replica of the vacuum. 
The number of such replicas depends on the parameters of the interaction, and we found
only one of them to exist for the combination of the linearly rising, Coulomb, and
constant potential with the realistic set of parameters, listed in Table~I. We tried
several contributions of the the parameters $\sigma_0$, $\alpha_s$ and $U$ consistent with
the phenomenology and numerically found it to be impossible to get rid of this replica. 

Thus starting from the system in its true vacuum state and exciting it, {\it e.g.}, 
by means of heating or placing a source of the strong field, one will finally
arrive at the totally disordered chirally symmetric phase, though at the intermediate
stage the system will quasistabilize (one or several times) that corresponds to its
reordering over a new, excited, vacuum state. It is no surprise that all excited
vacuum replicas are metastable since the hadronic spectra build over them contains 
tachyons. Still they may live for a finite period of time comparable with the
characteristic one of hadronic processes. Then the latter may go through the intermediate
stage with formation of the local bubble of the excited vacuum which decays then to the
ground-state true vacuum emitting pairs of photons and/or pions. The latter could be detected experimentally and,
thus, serve as a signal of the existence of the excited vacuum state. Moreover, such
measurements can serve as a direct probe of the QCD vacuum. 

Of course, the question may be raised, to what extent we rely on potential models of
QCD, and if the statement concerning the excited vacuum replicas is, indeed, model independent.
However potential models were shown to be very efficient in studies of
chiral symmetry breaking, hadronic properties and decays, and so on. Thus
we do believe that, at least qualitatively, their predictions, including the existence of excited 
vacuum replica, will persist in more realistic models for QCD, as well as in the true
theory, if it is ever solved.

\begin{acknowledgments}
The authors are grateful to Yu. S. Kalashnikova for drawing their attention to possible 
experimental tests where the excited vacuum could be observed, as well as for reading the
manuscript and critical comments. P. Bicudo thanks Anne-Christine Davis for discussions of the
linear potential, Dubravko Klabucar for discussions of the tunnel potential, and 
Felipe Llanes-Estrada and Steve Cotanch for discussions of the Coulomb potential.
A. Nefediev would like to thank the
staff of the Centro de F\'\i sica das Interac\c c\~oes Fundamentais (CFIF-IST) for cordial
hospitality during his stay in Lisbon and to acknowledge the financial support of 
RFFI grants 00-02-17836, 00-15-96786, and 01-02-06273,
INTAS-RFFI grant IR-97-232, and INTAS CALL 2000-110.
\end{acknowledgments}

\end{document}